\documentclass[aps,prl,superscriptaddress,twocolumn,floatfix,a4paper]{revtex4}

\usepackage{float}
\usepackage{amsmath}
\usepackage{amsfonts}
\usepackage{amssymb}
\usepackage{color}
\usepackage{units}
\usepackage{verbatim}
\usepackage{graphicx}
\usepackage{bbold}
\usepackage{soul}
\definecolor{DarkGreen}{rgb}{0,.5,0}
\definecolor{gray}{rgb}{.5,.5,.5}

\newcommand{\beq}{\begin{equation}}
\newcommand{\eeq}{\end{equation}}
\newcommand{\bea}{\begin{eqnarray}}
\newcommand{\eea}{\end{eqnarray}}
\newcommand{\nn}{\nonumber}

\newcommand{\smallfrac}[2]{\mbox{$\frac{#1}{#2}$}}
\newcommand{\bra}[1]{ \langle{#1} |}
\newcommand{\ket}[1]{ |{#1} \rangle}

\newcommand{\half}{\smallfrac{1}{2}}

\newcommand{\s}[1]{\hat\sigma_{#1}}

\newcommand{\blk}{\color{black}}

\newcommand{\expect}[1]{\langle{#1}\rangle}
\newcommand{\Bi}[1]{\mathcal{B}_{\mathit{#1}}}

\begin{document}

\title{Experimental demonstration of non-bilocal quantum correlations}

\author{Dylan J. Saunders$^\dag$}
\affiliation{Centre for Quantum Dynamics and Centre for Quantum Computation and
Communication Technology, Griffith University, Brisbane, 4111, Australia}
\affiliation{Clarendon Laboratory, University of Oxford, Parks Road, Oxford OX1 3PU, UK}
\author{Adam J. Bennet}
\affiliation{Centre for Quantum Dynamics and Centre for Quantum Computation and
Communication Technology, Griffith University, Brisbane, 4111, Australia}
\author{Cyril Branciard}
\affiliation{Institut N\'eel, CNRS and Universit\'e Grenoble Alpes, 38042 Grenoble Cedex 9, France}
\author{Geoff J. Pryde}
\affiliation{Centre for Quantum Dynamics and Centre for Quantum Computation and
Communication Technology, Griffith University, Brisbane, 4111, Australia}

\date{\today}

\begin{abstract}

Quantum mechanics admits correlations that cannot be explained by local realistic models. Those most studied are the standard local hidden variable models, which satisfy the well-known Bell inequalities. To date, most works have focused on bipartite entangled systems.
Here, we consider correlations between three parties connected via two independent entangled states. We investigate the new type of so-called ``bilocal'' models, which correspondingly involve two independent hidden variables. Such models describe scenarios that naturally arise in quantum networks, where several independent entanglement sources are employed. Using photonic qubits, we build such a linear three-node quantum network and demonstrate non-bilocal correlations by violating a Bell-like inequality tailored for bilocal models. Furthermore, we show that the demonstration of non-bilocality is more noise-tolerant than that of standard Bell non-locality in our three-party quantum network. 
\end{abstract}

\maketitle

Bell's theorem~\cite{BelPHY64} resolved the long-standing Einstein-Podolsky-Rosen debate~\cite{EinEtalPR35} by demonstrating that no local realistic theory can reproduce the correlations observed when performing appropriate measurements on some entangled quantum states -- so-called (Bell) non-local correlations~\cite{Brunner2014}. Entanglement now finds applications as a resource in many quantum information and communication protocols, as in e.g. Refs.~\cite{E91,Wooters1993}. In most fundamental or applied experiments to date, the entangled systems come directly from a single source. However, sometimes more than one source of entanglement is used, such as in protocols that rely on entanglement swapping~\cite{Zukowski93} to generate entanglement between two parties at the ends of a chain (even though they share no common history). Since the entanglement swapping results in a bipartite entangled state, one may examine this ``network'' scenario by considering only the non-locality of the correlations between the measurement outcomes at the terminal nodes. An ``event-ready'' Bell test~\cite{Zukowski93}, heralded on success signals from all intermediate nodes, would then aim to disprove a local theory that is based on a single local hidden variable (LHV) model. However, such a test ignores properties of the intermediate channel, such as the fact that the multiple sources of entanglement may be independent of each other. This raises an important fundamental question: how does source independence affect the notion of non-locality?

\begin{figure*}[t*]
\begin{center}
\includegraphics[width=.8\linewidth]{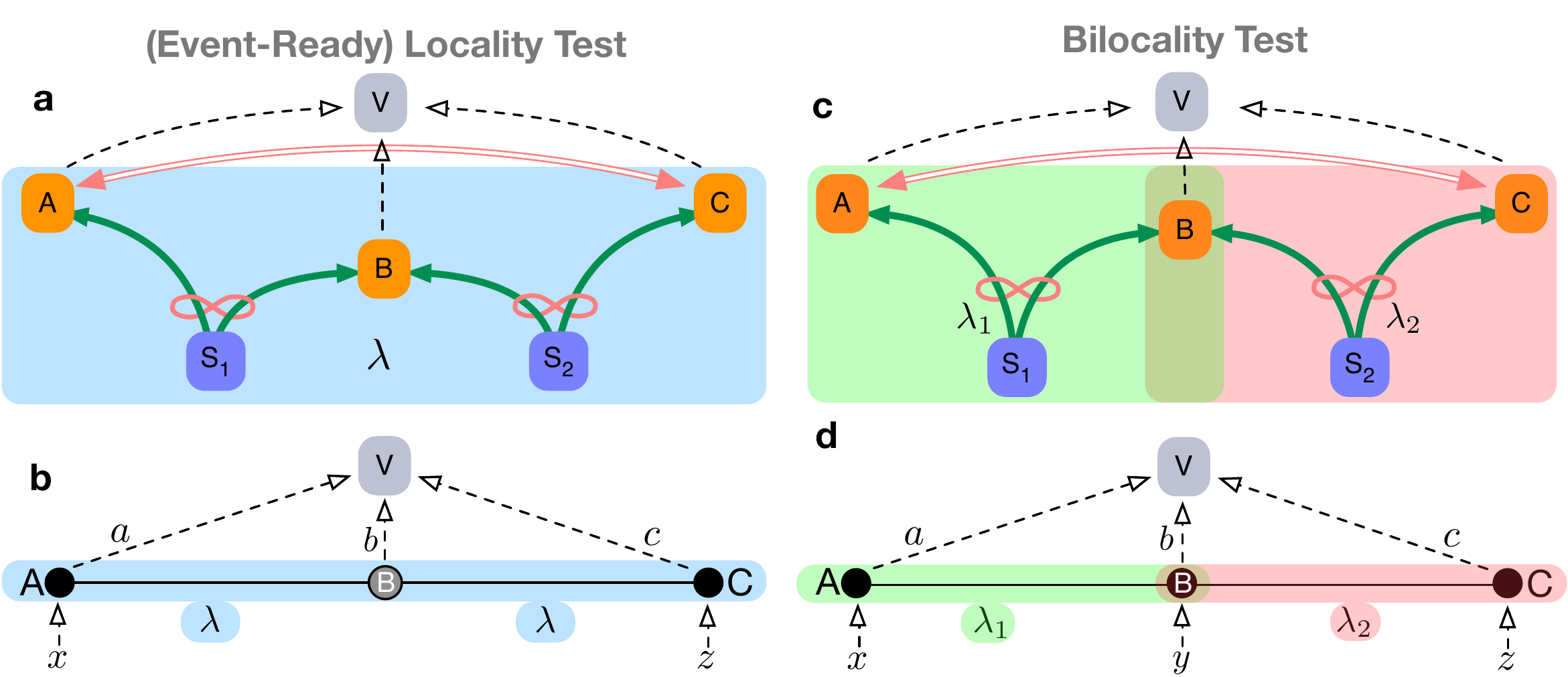}
\end{center}
\vspace{-3ex}
\caption{{\bf Tests of locality and bilocality (event-ready Bell test: a and b; bilocality: c and d).} Panels {\bf a} and {\bf c} show the experimental arrangement conceptually, involving: entangled photon pairs (green arrows) emitted from two independent sources ($S_1,S_2$); the three nodes: Alice (A), Bob (B) and Charlie (C); along with a referee Victor (V) who computes and analyses the correlations between the inputs and outputs ($x,(y),z,a,b,c$) of A, B, and C sent to him via classical communication channels (dashed arrows). The diagrams also show the regions of influence of the LHVs in the two models under consideration, $\lambda$ (blue shading) for the (Bell) locality case, or $\lambda_1$ (green shading) and $\lambda_2$ (salmon shading) for bilocality. The red double arrow represents the quantum correlations between the terminal nodes in each case. In the simplest event-ready implementation, {\bf a} and {\bf b}, Bob's measurement result $b$ is a binary variable that heralds a trial of a Bell test between Alice and Charlie, when Bob's (fixed) Bell state measurement successfully projects his two incoming systems onto e.g. the singlet state~\cite{Zukowski93}. In {\bf c} and {\bf d}, on the other hand, $b$ may be composed of more than 1 bit (corresponding to the result of a more informative joint measurement by Bob), and is taken into account in the test of a bilocal inequality. Panels {\bf b} and {\bf d} highlight the different network architecture of the two tests, including the nodes and connections (solid lines), the input measurement settings ($x,y,z$), and the measurement results ($a,b,c$).}
\label{fig:Q_network}
\end{figure*}

To address this question, a new type of LHV model was recently considered, where the independence properties of the different sources in an experimental setup are also imposed at the level of the hidden variables~\cite{Branciard2010,Branciard2012}. The simplest non-trival quantum network to analyse this new type of model is a three-node linear network, as depicted in Fig.~\ref{fig:Q_network}. In such a network, two independent entanglement sources connect the three nodes, Alice, Bob and Charlie; the corresponding model, that involves two independent LHVs, is termed ``bilocal''. Just like standard LHV models satisfy Bell inequalities, it was shown that bilocal models impose constraints on the corresponding correlations in the form of (nonlinear) Bell-like inequalities -- so-called ``bilocal inequalities'' -- which can be violated quantum mechanically~\cite{Branciard2010,Branciard2012}. One advantage of considering bilocal models is that one may demonstrate non-bilocality in situations where no non-locality could be obtained. For example, in an entanglement swapping experiment that generates a two-qubit Werner state between Alice and Charlie of the form $\rho_W(v) = v\,\ket{\psi}\!\bra{\psi}+(1{-}v)\,\smallfrac{\mathbb{1}}{4}$, where $\ket{\psi}$ is a maximally entangled state, and $\smallfrac{\mathbb{1}}{4}$ is the maximally mixed state, one requires a visibility $v > 1/\sqrt{2}$ to violate the commonly used Clauser-Horne-Shimony-Holt (CHSH) Bell inequality~\cite{CHSH:1969}, while bilocal inequalities can detect non-bilocality for any $v > 1/2$~\cite{Branciard2010,Branciard2012}; hence, one can certify the absence of a bilocal LHV model under more noise than for a Bell local model.

The aim of the present work is to investigate quantum non-bilocal correlations experimentally. We implement the scenarios of Fig.~\ref{fig:Q_network} in a photonic setup. In our experiment, the entangled photon pairs originate from two nonlinear crystals pumped separately, although by the same laser beam; to enhance the independence of the two sources, we actively destroy any coherence in the pump beam between the two crystals. We test two different bilocal inequalities, and find violations which allow us to disprove bilocal models for the quantum correlations we observe.

\emph{Local vs bilocal models.---}The differences between testing locality and bilocality on a three-node quantum network are highlighted in Fig.~\ref{fig:Q_network}. Let us first introduce a standard LHV three-party model: consider a tripartite probability distribution of the form
\beq P(a,b,c|x,y,z) = \int d\lambda ~\rho(\lambda) P(a|x, \lambda) P(b|y, \lambda) P(c|z, \lambda),\label{eq:3_bell_like}\eeq
where Alice, Bob, and Charlie have measurement inputs $x,y,z$, and measurement outputs $a,b,c$ respectively, and the LHV $\lambda$ with the distribution $\rho(\lambda)$ can be understood as describing the joint state of the three systems. $P(a|x, \lambda)$, $P(b|y, \lambda)$, and $P(c|z, \lambda)$ are the local probabilities for each separate outcome, given $\lambda$. A probability distribution $P(a,b,c|x,y,z)$ of the form of Eq.~\eqref{eq:3_bell_like} is said to be (Bell) local; one that cannot be expressed in that form is called (Bell) non-local~\cite{Brunner2014}. 

\begin{figure*}[t*]
\begin{center}
\includegraphics[width=.8\linewidth]{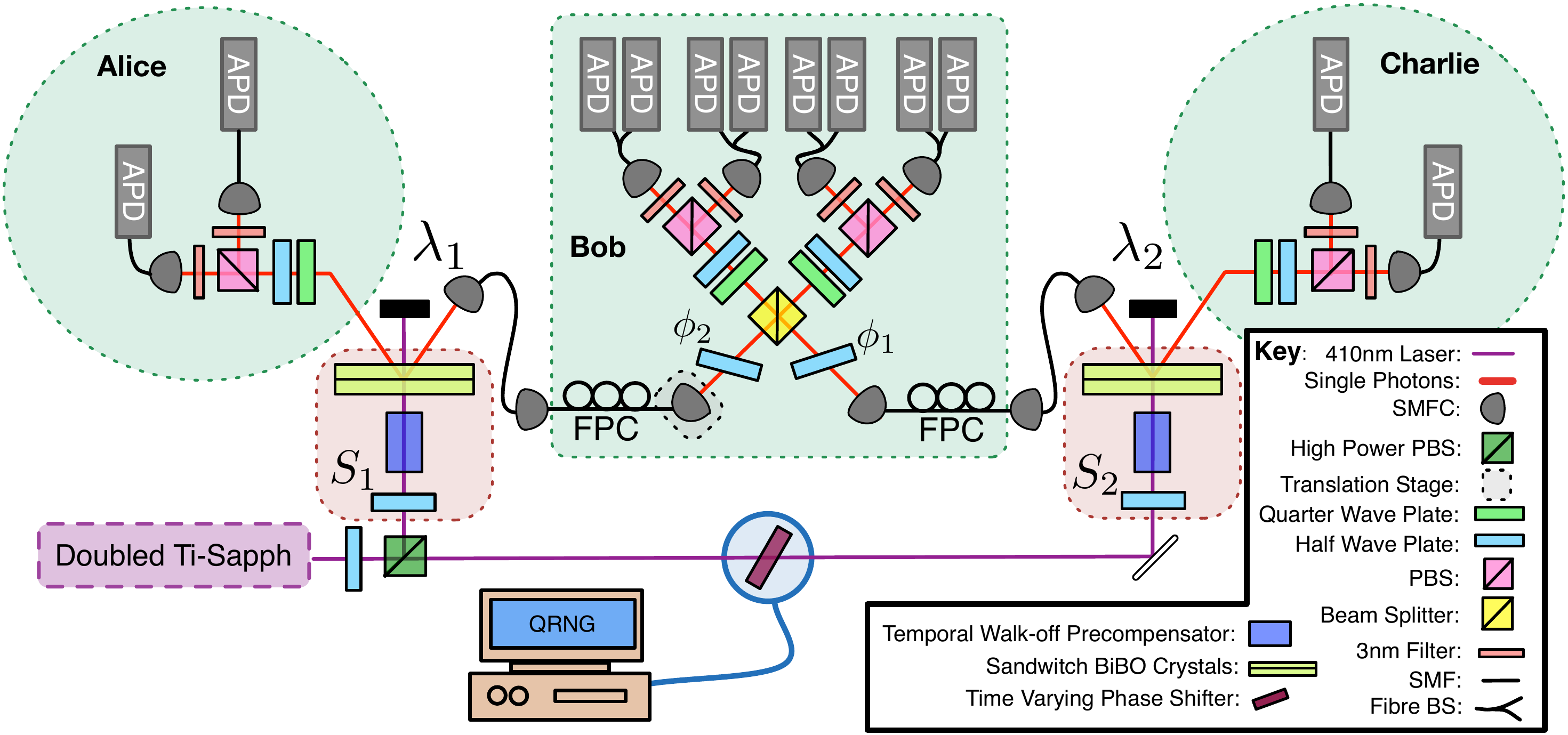}
\end{center}
\vspace{-5ex}
\caption{{\bf Experimental setup to test bilocality in a three-node quantum network.} The nodes -- Alice, Bob and Charlie -- are highlighted in green and the entanglement sources connecting them -- $S_1$ and $S_2$ -- in red. Both entanglement sources (sandwich BiBO crystals and  temporal walk-off precompensators~\cite{Kwiat}) are pumped by a mode-locked 410nm 80-MHz frequency-doubled titanium-sapphire oscillator. To substantiate the assumption that the LHVs $\lambda_1$ and $\lambda_2$ attached to the two sources are independent, we erase any coherence in the pump beam between $S_1$ and $S_2$ using a time-varying phase shifter set using a quantum random number generator (QRNG) -- see main text and Supplementary Material for details~\cite{Supp_Material}. Alice and Charlie implement their measurements (with settings $x,z$ and outputs $a,c$) using polarisation optics: quarter-wave plates, half-wave plates, polarising beam splitters (PBS), single mode fibres (SMFs) and single mode fibre couplers (SMFCs). Bob implements his Bell state measurement using a 50:50 beam slitter (BS) and polarisation optics. Bob ensures he implements the correct BSM (i.e., that he projects onto the desired Bell state in the simulated full BSM; see main text) by implementing single-qubit unitaries using a fibre polarisation controller (FPC) and phase gates ($\phi_1,\phi_2$: tilted half-wave plates). Bob also implements pseudo-number-resolving detectors on each of his four outputs, using a fibre 50:50 BS (Fibre BS) to split the output into two bucket avalanche photon detectors (APDs). We observe four-photon coincidence events -- one click for Alice and Charlie, and two for Bob -- on the APDs using a field programable gate array (FPGA) with a coincidence window of $3$ns to signify successful operation of our quantum network and to calculate all probabilities $P(a,b,c|x,z)$.}
\label{fig:exp}
\end{figure*}

In a practical experiment, where the above tripartite probability distribution is obtained by measuring some physical systems -- e.g. particles -- it is natural to assume that the LHV $\lambda$ originates from the source that prepares and sends those systems. For our three-node quantum network of Fig.~\ref{fig:Q_network} however, there are two independent sources of entangled particles -- $S_{1}$ and $S_{2}$. It is then natural to consider two local hidden variables, $\lambda_1$ and $\lambda_2$, one attached to each source, and write
\bea P(a,b,c|x,y,z) = \int d\lambda_{1~}d\lambda_{2}~ \rho(\lambda_{1},\lambda_{2}) P(a|x, \lambda_{1}) \nn\\ \times ~P(b|y,\lambda_{1},\lambda_{2}) P(c|z, \lambda_{2}). \label{eq:3_bell_like_2} \eea
Here, the local probabilities of each party are conditioned only on the LHV(s) attached to the source(s) they receive particles from: $\lambda_{1}$ for Alice, $\lambda_{2}$ for Charlie, and both $\lambda_{1}$ and $\lambda_{2}$ for Bob, at the intermediate node. So far, the correlations producible by the local decompositions in Eqs.~\eqref{eq:3_bell_like} and~\eqref{eq:3_bell_like_2} are equivalent. For example, the joint distribution of the two LHVs $\rho(\lambda_{1},\lambda_{2})$ could be non-zero only when $\lambda_{1}$ = $\lambda_{2}$ = $\lambda$~\cite{Branciard2010}. We shall however now introduce the critical \emph{bilocality assumption}, based on the physical arrangement of our quantum network: \emph{the independence of the two sources carries over to the local hidden variables $\lambda_{1}$ and $\lambda_{2}$}. That is, their joint distribution $\rho(\lambda_{1},\lambda_{2})$ must factorise: 
\beq\rho(\lambda_{1},\lambda_{2}) = \rho(\lambda_{1})\,\rho(\lambda_{2}).\label{eq:bilocal_assumption}\eeq 
\noindent Probability distributions $P(a,b,c|x,y,z)$ that can be expressed as Eq.~\eqref{eq:3_bell_like_2} with $\rho(\lambda_{1},\lambda_{2})$ satisfying Eq.~\eqref{eq:bilocal_assumption} are said to be ``bilocal''; those that cannot as termed ``non-bilocal''~\cite{Branciard2010,Branciard2012}.

\emph{Demonstrating non-bilocality.---} The decomposition of Eq.~\eqref{eq:3_bell_like_2}, together with~\eqref{eq:bilocal_assumption}, imposes certain restrictions on the correlations that can be produced by bilocal models. First note that any bilocal model is in particular Bell local, so that it must satisfy all Bell inequalities; any violation of a Bell inequality is already a demonstration of non-bilocality. However, it is also possible to derive stronger constraints for bilocal models, that specifically make use of the independence condition~\eqref{eq:bilocal_assumption}. In Ref.~\cite{Branciard2012} different bilocal inequalities were obtained, of the general form
\beq
\mathcal{B} := \sqrt{|I|} + \sqrt{|J|} \,\leq\, 1 \,, \label{eq:biloc_ineq}
\eeq 
where $I$ and $J$ are linear combinations of the observed probabilities $P(a,b,c|x,y,z)$ (see Supplementary Material for details~\cite{Supp_Material}). A violation of such an inequality, i.e. a $\mathcal{B}$ value greater than 1, is a proof of non-bilocality, as it rules out any possible bilocal model -- in a similar way that a CHSH value $\Bi{\text{CHSH}}$ greater than 2 disproves any Bell local model~\cite{CHSH:1969,Supp_Material}.

The bilocal inequalities above apply to scenarios where Alice and Charlie have binary inputs and outputs. As for Bob, we consider two cases that are of particular relevance experimentally. In the first case, he has a single fixed input and four possible outputs; following the notations of~\cite{Branciard2012}, we shall label this case $\mathit{14}$, and write the corresponding inequality as $\Bi{14} \leq 1$. In the second case, Bob still has a fixed input, but he now has three possible outputs; we shall label this case $\mathit{13}$, and write $\Bi{13} \leq 1$.
As discussed below, these two cases will correspond in the experiment to a full and a partial Bell state measurement, respectively.

\emph{Experiment.---} To test the two bilocal inequalities $\Bi{14},\Bi{13} \leq 1$, we realised a photonic implementation of an entanglement swapping type of experiment (e.g. ref.~\cite{Zeilinger1998}) implementing the three-node quantum network of Fig.~\ref{fig:Q_network}, see Fig.~\ref{fig:exp}. Two ``sandwich'' type-1 spontaneous parametric downconversion (SPDC) sources~\cite{Kwiat} supplied the entangled photonic links between the nodes. To justify that the bilocality assumption is reasonable, one should ideally have truly independent sources. In our case we used two separate nonlinear crystals to realise the parametric downconversion; however, the two crystals were pumped by a strong beam originating from the same laser. To increase the degree of independence between the two sources, we installed a time varying phase shifter (TVPS) in the pump beam before the source $S_2$. The TVPS comprised a rotatable optical flat connected to an automated stage and a remote quantum random number generator (QRNG)~\cite{anu}, adding a genuinely random phase offset between sources $S_1$ and $S_2$ on each trial of the experiment and thus destroying any quantum coherence (see Supplementary Material for details~\cite{Supp_Material}).

At the central node, Bob implements an entangling Bell state measurement (BSM)~\cite{Mattle96} to essentially fuse the two sources of entanglement $S_1$ and $S_2$ via entanglement swapping~\cite{Zukowski93}. Using linear optics only, it is impossible to construct an \textit{ideal} BSM device that reliably discriminates between all four Bell states, necessary for deterministic entanglement swapping~\cite{Suominen1999}. It is, however, possible to experimentally \textit{simulate} the statistics of an ideal BSM. We construct such a BSM device that projects onto one of the four Bell states. We then implement local unitaries to project separately, in different experimental runs, onto the three remaining states, and combine the statistics at the end of the experiment to mimic a universal BSM device. In this case, Bob's implemented measurement device has four input settings (one for each of the canonical Bell states $|\Phi^{+}\rangle$, $|\Phi^{-}\rangle$, $|\Psi^{+}\rangle$, and $|\Psi^{-}\rangle$) and one bit of output (indicating successful projection onto the relevant state). After recombining the statistics, Bob has simulated an ideal BSM device with a single input setting and four possible outputs, one corresponding to each of the four Bell states. It is precisely in this one-in/four-out scenario that one can test the $\Bi{14} \leq 1$ bilocal inequality introduced previously. Conveniently, it is also possible using linear optics to construct a \textit{partial} BSM device that projectively resolves two of the four Bell states (e.g. $|\Phi^{+}\rangle$ and $|\Phi^{-}\rangle$), accompanied by a third projection that groups the remaining two Bell states (e.g. $|\Psi^{\pm}\rangle$) into a single outcome~\cite{Kwiat1998} -- a single-input, three-output measurement allowing one to test the $\Bi{13} \leq 1$ bilocal inequality. As for Alice and Charlie, as mentioned above they should have binary inputs and outputs to test these two inequalities. We implemented projective measurements of the observables $\hat A_x$ and $\hat C_z$ (depending on the inputs $x,z = 0,1$) defined as $\hat{A_{0}}=\hat{C_{0}} = (\hat\sigma_{\textsc{z}}+\hat\sigma_{\textsc{x}})/\sqrt{2}$ and $\hat{A_{1}}=\hat{C_{1}} = (\hat\sigma_{\textsc{z}}-\hat\sigma_{\textsc{x}})/\sqrt{2}$ in the $\mathit{14}$ case (where $\hat\sigma_{\textsc{z,x}}$ are the standard Pauli matrices), and as $\hat A_0=\hat C_0=(\sqrt{2}\,\s{z}+\s{x})/\sqrt{3}$ and $\hat A_1=\hat C_1=(\sqrt{2}\,\s{z}-\s{x})/\sqrt{3}$ in the $\mathit{13}$ case, which in principle provide the optimal violations of the two inequalities~\cite{Branciard2012}.

Each entanglement source $S_i$ ($i=1,2$) ideally produces a pure Bell state. However, due to minor experimental imperfections, the produced states were close to Werner states (as described in the introduction) with visibility $v_i\gtrsim0.94$ (determined via quantum state tomography~\cite{White2007}) for both sources for all implementations of Bob's BSM. The fidelity of the Bell state measurement was maximised using single-mode fibres, narrowband frequency filters ($\sim3$nm full-width-half-maximum), and a high-precision translation stage, affording sub-coherence length timing resolution and ensuring high quality Hong-Ou-Mandel (HOM) interference: we measured a resultant HOM visibility of $v^{\textrm{BSM}}_{1,2}=(91\pm3)\%$ for $\mathit{13}$ and $v^{\textrm{BSM}}_{1,2}=(85\pm5)\%$ for $\mathit{14}$. The visibility of the closest Werner state to the resultant entangled state at Alice's and Charlie's terminal nodes (conditioned on Bob's BSM result) was estimated using quantum state tomography, yielding $v_{\mathit{13}}\approx0.85$ when Bob implements the $\mathit{13}$-BSM, and $v_{\mathit{14}}\approx0.78$ for Bob's $\mathit{14}$-BSM -- within error of the product of the visibility of each entangled source and the BSM visibility respectively, as expected. 

\begin{figure}[t!]
\begin{center}
\includegraphics[width=1\linewidth]{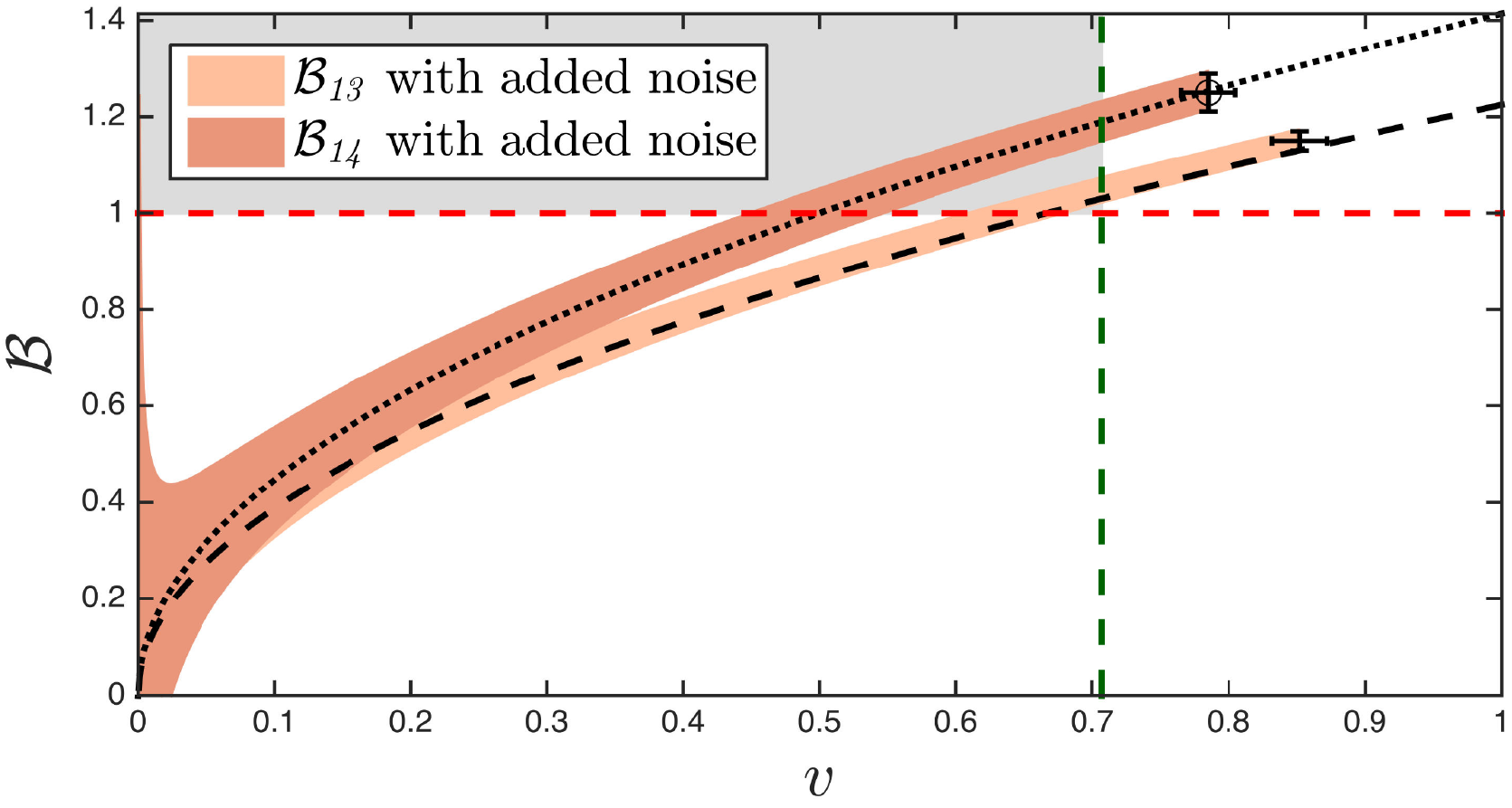}
\end{center}
\vspace{-5ex}
\caption{{\bf Evidence of noise-tolerant non-bilocality}. We measured bilocality parameters and estimated the corresponding visibilities $v$ for the best quality of our network that we obtained (the $\circ$ data point corresponds to $\Bi{14}$, and the other one to $\Bi{13}$). The error bars for $\Bi{}$ arise from Poissonian statistics, while the error on $v$ is calculated using the product of the source visibilities $v_i$ and the measured HOM dip visibility $v^{\textrm{BSM}}_{1,2}$ and agrees with the measured $\Bi{\text{CHSH}}$ (see main text and Supplementary Material for details~\cite{Supp_Material}). To test the noise-tolerance of non-bilocal correlations, we introduce noise in our data by randomly ``flipping'' trials of Alice's measurement~\cite{Supp_Material}, allowing us to predict the performance of our network to added white noise by simulating Werner states with $v_\mathit{13} \lesssim 0.85$ and $v_\mathit{14} \lesssim 0.78$, the maximum entanglement visibility of our networks. The orange shaded areas show the expected performance of our network under added noise to plus or minus one standard deviation. The dashed (dotted) lines are the expected values for $\Bi{14}(v)=\sqrt{2v}$ and $\Bi{13}(v)=\sqrt{3v/2}$~\cite{Branciard2012}. Both sets of values $\Bi{14}(v_\textrm{exp.})$ and $\Bi{13}(v_\textrm{exp.})$ occupy the grey shaded region that is non-bilocal ($\Bi{} > 1$, above red dashed line) and will not violate the CHSH inequality (for $v \leq 1/\sqrt{2}$, left of green dashed line) -- note that in our case with binary inputs and outputs for Alice and Charlie, and a fixed measurement setting for Bob, CHSH (with its symmetries) is the only relevant Bell inequality~\cite{Fine1982,Pironio2005}. This provides evidence for the higher noise-tolerance of non-bilocal correlations compared to Bell non-local correlations.}
\label{fig:13_noise}
\end{figure}

To further verify that our network was producing Werner-like states, we compared the measured CHSH inequality against the inferred visibilities above. We tested the CHSH inequality ($\Bi{\text{CHSH}} \leq 2$~\cite{CHSH:1969,Supp_Material}) on Alice's and Charlie's resultant state after successful entanglement swapping; a standard (event-ready~\cite{Zukowski93}) test of Bell locality. We recorded $\Bi{\text{CHSH}}^{\mathit{13}}=2.41\pm0.05$ and $\Bi{\text{CHSH}}^{\mathit{14}}=2.22\pm0.06$, agreeing with the measured $v$'s above (see Supplementary Details \cite{Supp_Material}), and both with clear violations of the bound. Next, the bilocal inequalities~\eqref{eq:biloc_ineq}, for both the full ($\mathit{14}$) and partial ($\mathit{13}$) Bell state measurements, were tested in our network, with clear violations in both cases: $\Bi{14}=1.25 \pm 0.04 > 1$ and $\Bi{13}=1.15 \pm 0.02 > 1$~\cite{Supp_Material}. To explore the noise robustness of our locality and bilocality tests, we add various amounts of white noise to our experimental data, see Fig.~\ref{fig:13_noise}. We implement this by swapping the labels on Alice's measurement outcomes on selected experimental runs, mimicking the effect of white noise by washing out the correlations (see Supplementary Material~\cite{Supp_Material}). This experimentally verified the prediction that in the presence of noise, there exists a region where non-bilocal correlations can be observed but non-local correlations cannot~\cite{Branciard2010,Branciard2012}.

\emph{Discussion.---} We have thus experimentally demonstrated the violation of two Bell-like inequalities tailored for quantum networks with independent entanglement sources, and verified that those inequalities can be violated at added noise levels for which a CHSH inequality cannot. As with quantum steering~\cite{Saunders2010}, for example, the addition of an extra assumption -- here, source independence -- relaxes the stringent intolerance to noise of non-locality demonstrations.

Our violation of bilocal inequalities shows in principle that no bilocal model can explain the correlations we observed. We acknowledge however that, like most Bell tests until very recently~\cite{Hensen2015,Giustina2015,Shalm2015}, our experiment is subject to some loopholes. In addition to a locality loophole (or sources are not space-like separated), and the detection loophole \cite{Pearle1970}, the specificity of the bilocality assumption opens a new ``source independence loophole'' when the entanglement sources are not guaranteed to be fully independent. In our experiment we enhanced the source independence, by erasing the quantum coherence between the pump beams of our two separate SPDC sources. Neverthess, the bilocality violations we observed could still in principle be explained by some hidden mechanism that would correlate the two sources (and the two LHVs $\lambda_1, \lambda_2$ attached to them in a bilocal model), for instance via the shared pump beam. In order to be able to draw more satisfying conclusions with regard to non-bilocality, the next step will be to realise a similar experiment with ``truly independent'' sources (following in the footsteps of Refs.~\cite{Rainer2009,Erven2014}) -- keeping in mind, however, that just like a Bell test can never rule out a superdeterministic explanation \cite{Bell1989}, it is impossible to guarantee that two separate sources are genuinely independent, as they could have been correlated at the birth of the universe.

The bilocality assumption, and its extension to ``$N-$locality'' in more complex scenarios involving $N$ independent sources, provides a natural framework to explore and characterize quantum correlations in multi-source, multi-party networks~\cite{Branciard2010,Branciard2012}. ``$N$-local inequalities'' have been derived in the line of Bell and bilocal inequalities~\cite{Fritz2012,Tavakoli2014,Hensen2014,Fritz2016,Lee2015,Chaves2016,Rosset2016,Tavakoli2016,Wolfe2016}, which could be tested in possible extensions of the present experiment, and in future larger quantum networks. An interesting question is, whether the violation of such inequalities could directly be exploited and could allow for useful applications in quantum information processing -- like the demonstration of Bell nonlocality or quantum steering can, e.g., be used to certify the security of quantum key distribution, or the privacy of randomness generation, in a device-independent way~\cite{Acin2007,Pironio2010,Colbeck2011,Branciard2012b}. We note that, contrary to the event-ready Bell test, the violation of the bilocal inequalities we tested here does not by itself certify that Bob must have performed an entangling measurement, and that Alice and Charlie end up sharing an entangled state (a counter-example is presented in the Supplementary Material~\cite{Supp_Material}) -- hence it is not sufficient for information processing protocols that require such a certification. However, we expect other possible applications to be discovered, that will fully harness the non-$N$-locality of quantum correlations, for instance in cases where non-locality cannot be demonstrated. The problem of characterizing and demonstrating non-$N$-local correlations will become more and more crucial as future quantum networks continue to grow in size and complexity.

\emph{Note added.---}During the preparation of our manuscript, we became aware of an independent experimental study of non-bilocality~\cite{Carvacho2016}. 

\emph{Acknowledgements.---}We thank Alastair Abbott, Nicolas Brunner, Nicolas Gisin, Stefano Pironio and Denis Rosset for helpful discussions. This research was conducted by the Australian Research Council Centre of Excellence for Quantum Computation and Communication Technology (Project number CE110001027). DJS and CB acknowledge EU Marie-Curie Fellowships PIIF-GA-2013-629229 and PIIF-GA-2013-623456 respectively. CB acknowledges a ``Retour Post-Doctorants'' grant from the French National Research Agency (ANR-13-PDOC-0026). The authors declare that they have no competing interests. All data needed to evaluate the conclusions in the paper are present in the paper and/or the Supplementary Material. Additional data related to this paper may be requested from the authors.


\clearpage 
\onecolumngrid

\begin{center}
\large{\bf Supplementary Material for:\\ Experimental demonstration of non-bilocal quantum correlations}
\end{center}

\setcounter{equation}{0}
\renewcommand{\theequation}{S\arabic{equation}}

\section*{Bilocal Inequalities}

The quantities $I$ and $J$ in the bilocal inequalities $\Bi{} := \sqrt{|I|} + \sqrt{|J|} \leq 1$ (Eq.~(4) of the main text) that we tested in our experiment are defined from the observed probabilities $P(a,b,c|x,z)$ as follows. (Please note, since we consider cases where Bob has a single fixed measurement setting $y$, we can simply ignore it when writing $P(a,b,c|x,z)$.) 
\medskip

Let us start with the full Bell state measurement (BSM), with 4 possible outcomes -- the case labelled $\mathit{14}$ (see main text). Here Bob's output consists of two bits, $b = b^0b^1$. Using some of the notations and forms introduced in Ref.~\cite{Branciard2012}, we first define, for $j = 0$ and $1$, the tripartite correlators (expectation values)
\beq
\expect{A_xB^j C_z}_{P_{\mathit{14}}}:=\sum_{a,b^0b^1,c} (-1)^{a+b^j+c} P_{\mathit{14}}(a,b^0b^1,c|x,z),
\eeq
where the sum is over all outputs $a,b^0,b^1,c = 0,1$ of the three parties.
These correlators, for the various values of $x,z=0,1$, then sum together in the following way to define $I_{\mathit{14}}$ and $J_{\mathit{14}}$ as
\beq
I_{\mathit{14}}:=\frac{1}{4}\sum_{x,z}\expect{A_xB^0 C_z}_{P_{\mathit{14}}}, \qquad
J_{\mathit{14}}:=\frac{1}{4}\sum_{x,z}(-1)^{x+z}\expect{A_xB^1 C_z}_{P_{\mathit{14}}}.
\eeq

\medskip

The case of a partial 3-outcome Bell state measurement, labelled $\mathit{13}$, is slightly complicated by the asymmetry in the partial BSM. Here we denote Bob's 3 possible outcomes as $b = b^0b^1 = 00,01, \{10~\text{or}~11\}$. The tripartite correlators are defined as
\beq
\expect{A_xB^0C_z}_{P_{\mathit{13}}}:=\sum_{a,c}(-1)^{a+c}\left[P_{\mathit{13}}(a,00,c|x,z)+P_{\mathit{13}}(a,01,c|x,z)-P_{\mathit{13}}(a,\{10 ~\text{or}~11\},c|x,z)\right]\eeq
and, restricting to the case where Bob gets one of the first two outcomes (i.e. $b^0=0$),
\beq
\expect{A_xB^1C_z}_{P_{\mathit{13}},b^0=0}:=\sum_{a,c}(-1)^{a+c}
\left[P_{\mathit{13}}(a,00,c|x,z)-P_{\mathit{13}}(a,01,c|x,z)\right].\eeq
Similarly as before, these correlators then sum together to now define
\beq
I_{\mathit{13}}:=\frac{1}{4}\sum_{x,z}\expect{A_xB^0 C_z}_{P_{\mathit{13}}},
\qquad
J_{\mathit{13}}:=\frac{1}{4}\sum_{x,z}(-1)^{x+z}\expect{A_xB^1 C_z}_{P_{\mathit{13}},b^0=0}.
\eeq

\medskip
In our experiment we realised both a (simulated) full and a partial BSM. In the first case, Bob's outputs $b = b^0b^1 = 00,01,10,11$ corresponded to the projections onto the Bell states $|\Phi^{+}\rangle$, $|\Phi^{-}\rangle$, $|\Psi^{+}\rangle$ and $|\Psi^{-}\rangle$, respectively. In the second case, $b = 00,01, \{10~\text{or}~11\}$ corresponded to $|\Phi^{+}\rangle$, $|\Phi^{-}\rangle$, and $|\Psi^{\pm}\rangle$ (which the partial BSM does not distinguish), respectively.
The following table shows, for both cases, the values of $I,J$ and $\mathcal{B}$ that we measured in our experiment.

\begin{center}
\begin{table}[h]
\begin{tabular}{c|c|c}
\hline
$I$ & $J$ & $\mathcal{B}$ \\
\hline
\hline
$I_{\mathit{14}}= 0.432 \pm 0.001$ & $J_{\mathit{14}}= 0.356 \pm 0.001$ & $\Bi{14}=1.25 \pm 0.04$ \\
$I_{\mathit{13}}= 0.6342 \pm 0.001$ & $J_{\mathit{13}}= 0.1252\pm 0.001$ & $\Bi{13}=1.15 \pm 0.02$ \\
\end{tabular}
\caption{\textbf{Observed violations of the inequalities $\Bi{} := \sqrt{|I|} + \sqrt{|J|} \leqslant 1$.} For comparison with the theoretical predictions, the optimal values for a perfect visibility ($v=1$, with perfect Bell states and measurements) are $I_{\mathit{14}}=J_{\mathit{14}} = \smallfrac{1}{2} =0.5$, $\Bi{\text{14}} = \sqrt{2} \simeq 1.414$, $I_{\mathit{13}}=\smallfrac{2}{3} \simeq 0.667$, $J_{\mathit{13}}=\smallfrac{1}{6} \simeq 0.167$ and $\Bi{\text{13}} = \sqrt{\smallfrac{3}{2}} \simeq 1.225$. For white noise characterised by a visibility $v$, the values of $I,J$ are simply multiplied by $v$, while the values of $\Bi{}$ are multiplied by $\sqrt{v}$~\cite{Branciard2012}.}
\end{table}
\end{center}

\medskip

For completeness, let us write explicitly the definition of the CHSH inequality~\cite{CHSH:1969} that we also tested in the experiment.
Here we define the bipartite correlators of Alice and Charlie as 
\beq
\expect{A_x C_z}_{|b=\checkmark\blk} := \sum_{a,c} (-1)^{a+c} P(a,c|x,z,b=\checkmark) \,
\eeq
where $P(a,c|x,z,b=\checkmark)$ is the joint probability distribution of Alice and Charlie's measurement outcomes, conditioned on their settings and on Bob's successful projection onto a fixed Bell state, $\ket{\psi^-}$ (`$b=\checkmark$') .
The CHSH inequality is then defined as
\beq
\Bi{\text{CHSH}} := \expect{A_0 C_0}_{|b=\checkmark\blk} + \expect{A_0 C_1}_{|b=\checkmark} - \expect{A_1 C_0}_{|b=\checkmark} + \expect{A_1 C_1}_{|b=\checkmark} \ \leq \ 2 \,.
\eeq

We tested this inequality using Alice and Charlie's measurement settings comprising of $\hat{A_{0}}=\sigma_{\textsc{z}}$, $\hat{A_{1}}=\sigma_{\textsc{x}}$, $\hat{C_{0}} = \frac{1}{\sqrt{2}}(\hat\sigma_{\textsc{z}}+\hat\sigma_{\textsc{x}})$ and $\hat{C_{1}} = \frac{1}{\sqrt{2}}(\hat\sigma_{\textsc{z}}-\hat\sigma_{\textsc{x}})$. As reported in the main text, we experimentally obtained a values of $\Bi{\text{CHSH}}^{\mathit{13}}=2.41\pm0.05$ and $\Bi{\text{CHSH}}^{\mathit{14}}=2.22\pm0.06$, indicating a clear violation of the CHSH inequality ($\Bi{\text{CHSH}}<2$), and thus demonstrating the non-locality of the correlations shared by Alice and Charlie at the end nodes of our network, conditioned on Bob's BSM result in our event-ready Bell test~\cite{Zukowski93}.

\section*{Characterisation of the Time Varying Phase Shift}

The Time Varying Phase Shift (TVPS -- a glass optical flat, Thorlabs part no. WG11050-B) used in the experiment served the purpose of erasing phase information in the pump beam between the sources $S_1$ and $S_2$, corroborating the independence condition. It was attached to a high resolution motorised rotation stage (Newport part no. URS100BCC) and a quantum random number generator (QRNG). The QRNG system used was the Australian National University QRNG system - a secure open source implementation~\cite{anu}. We accessed their live random number stream over the internet using a Matlab\textregistered~interface.

To characterise the TVPS, we constructed a Michelson-Morely interferometer and placed the TVPS into one arm of the interferometer. The TVPS was set to an initial testing angle of ~30$^{\circ}$ from normal, such that the slight rotations required to vary the global phase of source $S_2$ would cause minimal longitudinal disturbance to the beam pumping of $S_2$. The interference fringes of the Michelson-Morley interferometer were used to calibrate the TVPS, yielding 64 angular values (defined by the minimum resolution of the rotation stage, $0.005^{\circ}$) between $30^{\circ}$ and $30.315^{\circ}$ from normal, corresponding to phase shifts between $0$ and $2\pi$. After calibration, the TVPS was placed in the pump beam of Source 2 at an initial angle of $30^{\circ}$ from normal, mirroring the calibration. Our TVPS function took the QRNG randomly selected number, and mapped it into the range $[0,63]$, which then set the angle of the TVPS from a lookup table of angular values between $30^{\circ}$ and $30.315^{\circ}$. This corresponded to a quantum random phase between $0$ and $2\pi$ set every $\sim1$ms, or at a frequency $\sim1$KHz. This is much faster than the success rate of our quantum network ($\sim 0.1 $Hz), thus effectively erasing any relative phase information in the pump beam between $S_1$ and $S_2$ on the time-scale of the network. 

\section*{Photon Counting and Experimental Details}

Unwanted birefringence was compensated between $S_1, S_2$ and Bob using fibre polarisation controllers (FPCs) and phase-fixers. Successful entanglement swapping was post-selected using four-fold coincident detection. We simultaneously monitored the output of each polarisation analyser, involving 8 individual avalanche photo diodes (APDs) for the $\mathit{14}$ Bell state measurement (BSM -- see main text) and 12 APDs for the $\mathit{13}$ case (see Fig. 2 in the main paper) (Perkin Elmer SPCM-AQR-14-FC and custom single photon counting arrays). In the $\mathit{14}$ case, the fibre-BS are removed, and only single APDs are used at the outputs of Bob's polarisation projective measurements. In the latter case, Bob required number-resolving measurements. We implemented that by distinguishing two-photon events from single-photon events in the output modes of his BSM device. Since our APDs are not number resolving, we implemented pseudo-number-resolving-detectors using spatial multiplexing. Such a device, made of two APDs and a 50:50 beamsplitter, successfully detects two-photon pulses with $50\%$ probability. We corrected for this inefficiency in post-processing, allowing us to implement the $\mathit{13}$ BSM required to test the $\Bi{13}$ bilocal inequality. Furthermore, in all the inequalities tested we account for imbalances in detector efficiencies by swapping the outcomes of for inputs $x,y,z$ for half of the trials, effectively averaging over any bias in the outcomes for each party. We note, we make the fair-sampling assumption in the data presented in this work. 

\section*{Werner State Experimental Simulation}

In order to investigate the noise tolerant properties of testing different local and bilocal models in our quantum network, we add white noise to the to our data. Ideally, the joint state produced after entanglement swapping is one of the four Bell states. However, because of noise in such networks the states we produce can be approximated by a Werner state of the form
\beq
W=v\ket{\psi}\bra{\psi}+(1{-}v)\smallfrac{\mathbb{1}}{4},
\eeq
where $\ket{\psi}$ is the resulting shared Bell state after the swapping operation, and $\smallfrac{\mathbb{1}}{4}$ is the maximally mixed 2-qubit state. Our shared state between Alice and Charlie (outer nodes) had a visibility of $v_\mathit{13} \lesssim 0.85$ and $v_\mathit{14} \lesssim 0.78$ before introducing further white noise. This agrees with the source visibilities determined using quantum state tomography, and the visibility of the BSM determined by measuring a heralded HOM dip visibility, by scanning the automated delay stage in Bob's BSM apparatus. This also agrees with the measured CHSH parameter values for our network (see main manuscript). Unwanted higher order counts produced by the sources were suppressed by keeping the four-fold count rate low ($\sim 0.1 Hz$).

Our procedure was inspired by the effect of white noise on the observed statistics, by flipping Alice's measurements with probability $p=(1-v/2)$. This effectively simulated the effect of a depolarising channel for our polarisation-encoded qubits, allowing us to vary the value of $v$ for the Werner states produced in our network, up to the limit of $v \leq v_{\text{max}}$, where $v_{\text{max}}$ is the maximum entanglement quality of our network for both the $\mathit{13}$ and $\mathit{14}$ implementations.

\section*{Our Bilocal Inequalities Violations are Not Device-Independent Certifications of A-C Entanglement:\\
Counter-Example}

In order to get a Bell inequality violation in an event-ready Bell test based on a tripartite entanglement swapping scenario (see Fig.~1 of the main text), a strict requirement is that Bob performs an entangling measurement, so that Alice and Charlie's particles end up being entangled. The violation of a Bell inequality between Alice and Charlie (conditioned on Bob's output) thus witnesses, in a device-independent way, the entanglement between Alice and Charlie.
One may then wonder if this property also holds for the violation of the two bilocal inequalities of the form $\Bi{} := \sqrt{|I|} + \sqrt{|J|} \leq 1$ that we tested in our experiment. The answer is negative, as shown by the following counter-example.

\medskip

Consider a case where the source $S_1$ sends the pure maximally entangled state $\rho_{AB} = \ket{\Phi^+}\!\bra{\Phi^+}$ to Alice and Bob, and the source $S_2$ sends the separable, correlated mixed state $\rho_{BC} = \half \ket{00}\!\bra{00} + \half \ket{11}\!\bra{11}$ to Bob and Charlie.
Take Alice and Charlie's measurements to be $\hat{A_{0}} = \frac{1}{\sqrt{2}}(\hat\sigma_{\textsc{z}}+\hat\sigma_{\textsc{x}})$, $\hat{A_{0}} = \frac{1}{\sqrt{2}}(\hat\sigma_{\textsc{z}}-\hat\sigma_{\textsc{x}})$, $\hat{C_{0}} = \mathbb{1}$ and $\hat{C_{1}} = \hat\sigma_{\textsc{z}}$.
Bob's measurement is described as follows: he first measures the qubit received from the source $S_2$ in the computational basis $\{\ket{0},\ket{1}\}$. If he gets the result $0$, then he defines $b^0$ to be the result of a $\hat\sigma_{\textsc{z}}$ measurement on the qubit received from $S_1$, and attributes a random value to $b^1$; if he gets the result $1$, then he defines $b^1$ to be the result of a $\hat\sigma_{\textsc{x}}$ measurement on the qubit received from $S_1$, and attributes a random value to $b^0$. In the $\mathit{14}$ case Bob then outputs both bits $b^0$ and $b^1$; in the $\mathit{13}$ case he simply groups the outputs $b^0b^1 = \{10~\text{or}~11\}$.

Clearly, Bob's measurement is separable (he never projects the two qubits he receives from the two sources onto an entangled state). Also, the state shared by Bob and Charlie is separable. Hence, no entanglement is shared between Alice and Charlie at any point of the protocol, whatever Bob's measurement outcomes.
The quantum mechanical predictions for the values of $I$ and $J$ are found to be $I_{\mathit{14}}=J_{\mathit{14}} = I_{\mathit{13}} = \frac{1}{2\sqrt{2}}$ and $J_{\mathit{13}} = \frac{1}{4\sqrt{2}}$, giving $\Bi{14} = 2^{1/4} > 1$ and $\Bi{13} = \frac{\sqrt{2}+1}{2^{5/4}} > 1$.

\medskip

This counter-example shows that one can obtain indeed a violation of the two bilocal inequalities above even when Bob performs a non-entangling joint measurement, and Alice and Charlie's particles never get entangled; in that case the violation is only due to the initial entanglement between Alice and Bob. This is in fact not too surprising. Indeed, for any bilocal tripartite correlation of the form of Eq.~(2) in the main text, Alice and Bob's bipartite marginal correlation is necessarily Bell local. Hence, whatever the correlations with Charlie, as soon as Alice and Bob's correlations are non-local, then the tripartite correlations will be non-bilocal. It would however be interesting to find bilocal inequalities whose violations do certify entanglement between the end nodes of the network, and which would not simply reduce to standard Bell inequalities (as in the event-ready Bell test mentioned above).
 
\end{document}